\documentstyle[epsfig,preprint,eqsecnum,aps]{revtex}
\tightenlines
\begin{document}
% repeat the \author\address pair as needed
\newcommand{\beq}{\begin{equation}}
\newcommand{\eeq}{\end{equation}}
\newcommand{\bea}{\begin{eqnarray}}
\newcommand{\eea}{\end{eqnarray}}
\newcommand{\ek}{\not{\varepsilon}}
\newcommand{\eek}{\vec{\varepsilon}}
\newcommand{\gp}{ {\cal G\/}_{F} }
% \draft command makes pacs numbers print

\draft
\preprint{BETHLEHEM-Phys-HEP980501}
\title{Deconfinement in the Quark Meson Coupling Model}
\author{I. Zakout and H. R. Jaqaman}
\address{Department of Physics, Bethlehem University,
P.O. Box 9, Bethlehem, Palestine}
\maketitle
\begin{abstract}
The Quark Meson Coupling Model which describes nuclear matter as
a collection of non-overlapping MIT bags interacting 
by the self-consistent exchange of scalar and vector mesons 
is used to study nuclear matter at finite temperature. 
In its modified version,
the density dependence of the bag constant  is introduced by  
a direct coupling between the bag constant and the scalar mean field.
In the present work,  the coupling of the scalar mean field with the
constituent quarks is considered exactly through the solution of 
the Dirac equation.
Our results show that a phase transition takes place  at a critical
temperature
around 200 MeV in which  the scalar mean field takes a nonzero value
at zero baryon density.
Furthermore it is found that the bag constant decreases significantly when 
the temperature increases above this critical temperature indicating 
the onset of quark deconfinement.

\end{abstract}
\narrowtext
%
% insert suggested PACS numbers in braces on next line

\section{Introduction}

The Quark Meson Coupling Model (QMC), initially proposed by 
Guichon\cite{Guichon},
describes nuclear matter as a collection of non-overlapping MIT bags
interacting by the self-consistent exchange
of scalar $\sigma$ and vector $\omega$ mesons in the mean field
approximation\cite{Saito,Jin96,Jin96b}.
It is a simple extension of the Walecka model
\cite{Walecka,SerotA,Furnstahl} except that the meson fields are 
coupled directly to the constituent quarks themselves rather 
than to the nucleons as in the Walecka model. The QMC model
thus incorporates explicitly the quark degrees of freedom.
This simple model was  later refined by including the nucleonic 
fermi motion and the center-of-mass correction
\cite{Fleck} to the bag energy
and applied to a variety of problems\cite{SaitoII,CohenW}.

In a modification of the original QMC model it has been suggested  
by Jin and Jennings\cite{Jin96,Jin96b} that a reduction of 
the bag constant in nuclear matter relative to 
its free-space value may be essential for the success of relativistic 
nuclear phenomenology and that it may play an important role 
in low and medium nuclear physics such as understanding 
the EMC effect\cite{EMC}.
The density-dependence of the bag constant is introduced 
in the present work by coupling 
it to the scalar meson field as suggested in Ref.\cite{Jin96}.
It was found by Jin and Jennings that when 
the bag constant 
is significantly reduced in nuclear medium with respect to 
its free-space value, large cancelling isoscalar Lorentz scalar 
and vector potentials for the nucleon in nuclear matter 
emerge naturally.
Such potentials are comparable to those suggested by relativistic
nuclear phenomenology and finite density QCD sum rules
\cite{CohenQCD}.

Recently, Panda {\em et al.}\cite{Panda} 
have studied nuclear matter at finite 
temperature using this  modified version of the QMC model.
They determined the scalar mean field by minimizing the grand thermodynamical 
potential with respect to the $\sigma$ field and using the self-consistency
condition which relates  the vector mean field to the baryon density. 
They found that the nucleon properties at finite temperature and/or 
nonzero baryon density are appreciably different
from  their  zero temperature vacuum values.

Our present work is essentially an extension of the work 
of Panda {\em et al.}\cite{Panda}.
We intend to  use the QMC model at finite temperature
and to take the medium dependence of the bag parameters into account.
We however will attempt to solve the self-consistency condition 
for the $\sigma$ field exactly by taking into consideration 
the full coupling of the scalar mean field to the internal quark 
structure by means of the solution of the point-like
Dirac equation with the required boundary condition of  confinement 
at the surface of the bag as suggested by Refs.\cite{Saito,Jin96}.
This was not done exactly  in the solution of the self-consistency
condition for the $\sigma$ field by Panda et al.

The outline of the paper is as follows. In section II, 
we present the QMC model for nuclear matter at finite temperature, 
together with the details of the  self-consistency condition 
for the scalar mean field. 
In section III, we  discuss our results and present our conclusions.

\section{QMC model for nuclear matter at finite temperature}
The QMC model at finite temperature is described in detail in Ref.\cite{Panda}.
We give here the essential equations necessary for the present 
calculations. 
The quark field $\psi_{q}(\vec{r},t)$ inside the bag satisfies
the Dirac equation
\begin{eqnarray}
\left[ i\gamma^{\mu}\partial_{\mu}-(m_{q}^{0}-g_{\sigma}^{q}\sigma)
-g_{\omega}^{q}\omega\beta\right]\psi_{q}(\vec{r},t)=0.
\label{Dirac} %(2.1)
\end{eqnarray}
Here the single-particle quark and antiquark energies in units of
$R^{-1}$ are given as
\begin{eqnarray}
\epsilon^{n\kappa}_\pm=\Omega^{n\kappa}\pm g_{\omega}^{q}\omega R
\label{EPN} %(2.2)
\end{eqnarray}
where
\begin{eqnarray}
\Omega^{n\kappa}=\sqrt{ x^{2}_{n\kappa} + R^{2}{m^{*}}^{2}_{q} }
\label{Omegnk} %(2.3)
\end{eqnarray}
and $m^{*}_{q}=m^{0}_{q}-g^{q}_{\sigma}\sigma$ is the effective
quark mass.
The boundary condition at the bag surface is given by
\begin{eqnarray}
i\gamma\cdot \hat{n} \psi_{q}^{n\kappa}=\psi_{q}^{n\kappa},
\label{roots} % (2.4)
\end{eqnarray}
which determines the quark momentum $x_{n\kappa}$ in the state
characterized by specific values of $n$ and $\kappa$.
For given values of the bag radius $R$ and the scalar field
$\sigma$, the quark momentum $x_{n\kappa}$ is obtained from
Eq.(\ref{roots}).
The quark chemical potential $\mu_{q}$ assuming that there are
three quarks in the nucleon bag is determined through
\begin{eqnarray}
n_{q}&=&3 \nonumber \\
     &=&3\sum_{n\kappa}\left[
 \frac{1}{e^{(\epsilon^{n\kappa}_{+}/R-\mu_{q})/T}+1}
-\frac{1}{e^{(\epsilon^{n\kappa}_{-}/R+\mu_{q})/T}+1}
\right].
\label{nq} % (2.5)
\end{eqnarray}
The total energy from the quark and antiquark is
\begin{eqnarray}
E_{tot}=3\sum_{n\kappa}
\frac{\Omega^{n\kappa}}{R}\left[
 \frac{1}{e^{(\epsilon^{n\kappa}_{+}/R-\mu_{q})/T}+1}
+\frac{1}{e^{(\epsilon^{n\kappa}_{-}/R+\mu_{q})/T}+1}
\right].
\label{Etot} % (2.6)
\end{eqnarray}
The bag energy is given by 
\begin{eqnarray}
E_{bag}=E_{tot}-\frac{Z}{R}+\frac{4\pi}{3}R^{3}B(\sigma).
\label{Ebag} % (2.7)
\end{eqnarray}
where $B(\sigma)$ is the bag parameter.
In the simple QMC model, the bag parameter $B$ is taken as
$B_{0}$ corresponding to its value for a free nucleon.
The medium effects are taken into account in the modified 
QMC model by the following ansatz for the bag parameter
\cite{Panda,Jin96}
\begin{eqnarray}
B=B_{0}\exp\left(-\frac{4g^{B}_{\sigma}\sigma}{M_{N}}\right)
\label{BB0} % (2.8)
\end{eqnarray}
with $g^{B}_{\sigma}$ as an additional parameter.       
The spurious center-of-mass momentum in the bag is subtracted
to obtain the effective nucleon mass\cite{Fleck}

\begin{eqnarray}
M^{*}_{N}=\sqrt{E^{2}_{bag}-<p^{2}_{cm}>}
\label{MNSTAR} % (2.9)
\end{eqnarray}
where
\begin{eqnarray}
<p^{2}_{cm}>=\frac{<x^{2}>}{R^{2}}
\label{PCM} % (2.10)
\end{eqnarray}
and
\begin{eqnarray}
<x^{2}>=3\sum_{n\kappa} x^{2}_{n\kappa}
\left[
 \frac{1}{e^{(\epsilon^{n\kappa}_{+}/R-\mu_{q})/T}+1}
+\frac{1}{e^{(\epsilon^{n\kappa}_{-}/R+\mu_{q})/T}+1}
\right].
\label{x2} % (2.11)
\end{eqnarray}
The bag radius $R$ is obtained through the minimization of 
the nucleon mass with respect to the bag radius 
\begin{eqnarray}
\frac{\partial M^{*}_{N}}{\partial R}=0.
\label{MNR} % (2.12)
\end{eqnarray}
%%%%%%%%%%%%%%%%%%%%%
%%%%%%%%%%%%%%%%%%%%%
The total energy density at finite temperature $T$
and at finite baryon density $\rho_{B}$ is
\begin{eqnarray}
\epsilon=
\frac{\gamma}{(2\pi)^{3}}\int
d^{3}k\sqrt{k^{2}+{M_{N}^{*}}^{2}}(f_{B}+\overline{f}_{B})
+\frac{g^{2}_{\omega}}{2m^{2}_{\omega}}\rho^{2}_{B}
+\frac{1}{2}m^{2}_{\sigma}\sigma^{2},
\label{density} % (2.13)
\end{eqnarray}
where $\gamma=4$ is the spin-isospin degeneracy factor and $f_{B}$ and
$\overline{f}_{B}$ are the Fermi-Dirac  distribution
functions for the baryons and antibaryons
\begin{eqnarray}
f_{B}=\frac{1}{e^{(\epsilon^{*}-\mu^{*}_{B})/T}+1},
\label{fB} % (2.14)
\end{eqnarray}
and
\begin{eqnarray}
\overline{f}_{B}=\frac{1}{e^{(\epsilon^{*}+\mu^{*}_{B})/T}+1},
\label{fBAR} % (2.15)
\end{eqnarray}
with $\epsilon^{*}=\sqrt{ k^{2}+{M^{*}_{N}}^{2} }$
the effective nucleon energy and $\mu^{*}_{B}=\mu-g_{\omega}\omega$
the effective baryon chemical potential.
The chemical potential for a given density $\rho_{B}$ is determined by
\begin{eqnarray}
\rho_{B}=\frac{\gamma}{(2\pi)^{3}}\int d^{3}k(f_{B}-\overline{f}_{B})
\label{rhoB} % (2.16)
\end{eqnarray} 
where 
\begin{eqnarray}
\omega=\frac{g_{\omega}}{m^{2}_{\omega}}\rho_{B}.
\label{omegrho} % (2.17)
\end{eqnarray}
The pressure is the negative of the grand thermodynamic potential 
and is given by

\begin{eqnarray}
P=
\frac{1}{3}\frac{\gamma}{(2\pi)^{3}}\int d^{3} k
\frac{k^{2}}{\epsilon^{*}}(f_{B}+\overline{f}_{B})
+\frac{1}{2}m^{2}_{\omega}\omega^{2}
-\frac{1}{2}m^{2}_{\sigma}\sigma^{2}.
\label{pressure} % (2.20)
\end{eqnarray}

The scalar mean field $\sigma$ is determined through the 
minimization of the thermodynamic potential or the maximizing 
of the pressure  $\frac{\partial P}{\partial \sigma}=0$ \cite{Panda}.
The pressure depends explicitly on the scalar mean field $\sigma$ 
through the last term in Eq.(\ref{pressure}). It also depends 
on the nucleon effective mass $M^{*}_{N}$
which in turn also depends on  $\sigma$. 
If we write the pressure  as a function of $M_{N}^{*}$ and $\sigma$
\cite{Jin96,Saito},
the extremization of  $P(M^{*}_{N},\sigma)$ 
with respect to the scalar mean field $\sigma$ can be written as 
\begin{eqnarray}
\frac{\partial P}{\partial \sigma}=
\left( \frac{\partial P}{\partial M^{*}_{N}} \right)_{\mu_{B},T}
\frac{\partial M^{*}_{N}}{\partial \sigma}
+\left(\frac{\partial P}{\partial \sigma}\right)_{M^{*}_{N}}=0,
\label{preseg}
\end{eqnarray}
where
\begin{eqnarray}
\left(\frac{\partial P}{\partial \sigma}\right)_{M^{*}_{N}}=
-m^{2}_{\sigma}\sigma,
\label{PSE}
\end{eqnarray}
and
\begin{eqnarray}
\left( \frac{\partial P}{\partial M^{*}_{N}} \right)_{\mu_{B},T}=
&-&\frac{\gamma}{3}\frac{1}{(2\pi)^{3}}
\int d^{3} k
\frac{k^{2}}{{\epsilon^{*}}^{2}}\frac{M^{*}_{N}}{\epsilon^{*}}
\left[f_{B}+\overline{f}_{B}\right]
\nonumber \\
&-&\frac{\gamma}{3}\frac{1}{(2\pi)^{3}}
\frac{1}{T}\int d^{3} k
\frac{k^{2}}{\epsilon^{*}}\frac{M^{*}_{N}}{\epsilon^{*}}
\left[f_{B}(1-f_{B})+\overline{f}_{B}(1-\overline{f}_{B})\right]
\nonumber \\
&-&\frac{\gamma}{3}\frac{1}{(2\pi)^{3}}
\frac{1}{T}g_{\omega}
\left(\frac{\partial \omega}{\partial M^{*}_{N}}\right)_{\mu_{B},T}
\int d^{3} k
\frac{k^{2}}{\epsilon^{*}}
\left[f_{B}(1-f_{B})-\overline{f}_{B}(1-\overline{f}_{B})\right]
\nonumber \\
&+&m^{2}_{\omega}\omega 
\left(\frac{\partial \omega}{\partial M^{*}_{N}}\right)_{\mu_{B},T}.
\label{PME}
\end{eqnarray}
Since the baryon chemical potential $\mu_{B}$ and temperature are
treated as input parameters, the variation of the vector mean field
$\omega$ with respect to the effective nucleon mass $M^{*}_{N}$
at a given value of the baryon density $\rho_{B}$ reads
\begin{eqnarray}
\left(\frac{\partial \omega}{\partial M_{N}^{*}}\right)_{\mu_{B},T}=
-\frac{
\frac{g_{\omega}}{m^{2}_{\omega}}
\frac{\gamma}{(2\pi)^{3}}\int d^{3}k
\frac{M_{N}^{*}}{\epsilon^{*}}
\left[f_{B}(1-f_{B})-\overline{f}_{B}(1-\overline{f}_{B})\right]
}
{1+
\frac{g^{2}_{\omega}}{m^{2}_{\omega}}
\frac{\gamma}{(2\pi)^{3}}
\int d^{3}k
\left[f_{B}(1-f_{B})+\overline{f}_{B}(1-\overline{f}_{B})\right]
}.
\label{OME}
\end{eqnarray}
The coupling of the scalar mean field $\sigma$ with the constituent 
quark in the non-overlapping MIT bag through the solution 
of the point like Dirac equation should be taken into account 
to satisfy the self-consistency condition.
This constraint is essential to obtain the correct solution
of the scalar mean field $\sigma$.
The differentiation  of the effective nucleon mass $M_{N}^{*}$
with respect to $\sigma$ gives 
\begin{eqnarray}
\frac{\partial M^{*}_{N}}{\partial \sigma}=
\frac{
E_{bag}\frac{\partial E_{bag}}{\partial \sigma}
-\frac{1}{2}\frac{1}{R^{2}}\frac{\partial <x^{2}>}{\partial \sigma}
}{M^{*}_{N}},
\label{MSE}
\end{eqnarray}
where
\begin{eqnarray}
\frac{\partial E_{bag}}{\partial \sigma}=
\sum_{n\kappa} \frac{\partial E_{bag}}{\partial \Omega^{n\kappa}_{q}}
\frac{\partial \Omega^{n\kappa}_{q}}{\partial \sigma}
+ \left(\frac{\partial E_{bag}}{\partial \sigma}\right)%
_{\{\Omega^{n\kappa}_{q}\}},
\label{ESE}
\end{eqnarray}
and
\begin{eqnarray}
\frac{\partial <x^{2}>}{\partial \sigma}=
\sum_{n\kappa} \frac{\partial <x^{2}>}{\partial \Omega^{n\kappa}_{q}} 
\frac{\partial \Omega^{n\kappa}_{q}}{\partial \sigma}
+\left( \frac{\partial <x^{2}>}{\partial \sigma}\right)%
_{\{\Omega^{n\kappa}_{q}\}}.
\label{XSE}
\end{eqnarray}
The $\frac{\partial \Omega^{*}_{q}}{\partial \sigma}$ depends
on $x_{n\kappa}$. Its evaluation can be obtained from the solutions 
of the point like Dirac equation for the constituent quarks 
which satisfy the required boundary condition 
on the surface of the bag\cite{Saito,Jin96}.
In our case it reads\cite{Saito,Jin96}
\begin{eqnarray}
\left(\frac{\partial \Omega^{n\kappa}_{q}}{\partial \sigma}\right)
=-g^{q}_{\sigma}R
<\overline{\psi}^{n\kappa}|\psi^{n\kappa}>.
\end{eqnarray}

\section{Results and Discussions}

We have studied nuclear matter at finite temperature
using the modified quark meson coupling model 
which takes the medium-dependence 
of the bag into account. 
We choose a direct coupling  of the bag constant 
to the scalar mean field $\sigma$ in the form given 
in Eq.(\ref{BB0}).
The bag  parameters are taken as those adopted by
Jin and Jennings\cite{Jin96} 
where $B^{1/4}_{0}=188.1$ MeV and $Z_{0}=2.03$  
are chosen to reproduce the free nucleon mass 
$M_{N}$ at its experimental 
value 939 MeV and bag radius $R_{0}=0.60$ fm. 
The current quark mass $m_{q}$ is taken equal to zero.
For $g_{\sigma}^{q}=1$, the values of the vector meson coupling 
and the parameter $g_{\sigma}^{B}$ as fitted from the saturation 
properties of nuclear  matter, are given 
as $g^{2}_{\omega}/2\pi=5.24$ 
and ${g^{B}_{\sigma}}^{2}/4\pi$=3.69.

We first solve Eqs.(\ref{rhoB}) and (\ref{omegrho})
for given values of temperature and density $\rho_{B}$ 
to determine the baryon chemical potential $\mu_{B}$. 
This constraint is given in terms of the effective nucleon mass 
$M^{*}_{N}$ which depends  on the bag radius R, 
the quark chemical potential $\mu_{q}$ 
and the mean field $\sigma$.
For given values of $\sigma$ and $\omega$ the bag radius and the
quark chemical potential $\mu_{q}$ are obtained using the 
self-consistency
conditions Eq.(\ref{MNR}) and Eq.(\ref{nq}), respectively.
The pressure is evaluated 
for specific values  of temperature
and $\mu_{B}$ which now becomes an input parameter. 
We then determine the value of $\sigma$ by using 
the maximization condition given in Eq.(\ref{preseg}).
This method is analogous to the method used by Saito\cite{Saito}
and Jin\cite{Jin96} for the zero temperature case.
It takes into account the coupling of the constituent quarks 
with the scalar mean field in the frame of the solution of 
the point like Dirac equation exactly. 
It differs from the minimization method used in Ref.\cite{Panda}.

Fig.1, displays various pressure isotherms vs. the baryon density.
The pressure has the usual trend of increasing with temperature 
and density. 
It is important to note that the pressure attains a nonzero value 
at zero baryon density above a critical temperature  
$T_{c}\simeq 200$ MeV. 
This occurs because the scalar mean field $\sigma$ attains 
a nonzero value at zero baryon density $\rho_{B}=0$
[see discussion concerning Fig.2 below] as was also observed
in the Walecka\cite{Furnstahl} model where it leads 
to a sharp fall in the effective nucleon mass at $T_{c}$.
This rapid fall of $M_{N}^{*}$ with increasing 
temperature resembles a phase transition when the system 
becomes a dilute gas of baryons in a sea of mesons and baryon-antibaryon 
pairs.

In Fig.2, we display the scalar mean field $\sigma$ as 
a function of the baryon density for various temperatures.
Fig.2 (a) indicates that the value of $\sigma$ initially  decreases 
with increasing temperature for temperatures less than 200 MeV. 
However, by the time the temperature reaches 150 MeV there are indications of 
an increase in $\sigma$ at very low baryon densities with a nonzero value 
at $\rho_{B}=0$. 
For still higher temperatures, as can be seen in Fig.2 (b), 
the situation is more dramatic with the value of $\sigma$  increasing  
with temperature for all values of $\rho_{B}$. This is an indication of 
a phase transition to a system of baryon-antibaryon pairs 
at very low densities as mentioned above.  

The density and temperature dependence of the baryon effective 
mass $M_{N}^{*}$ is shown in Fig 3. 
For low baryon density $\rho_{B}$, as the temperature is increased, $M_{N}^{*}$  
first increases slightly and then decreases rapidly for $T=200$ MeV 
for densities less than about 0.2 fm$^{-3}$.
This rapid decrease of $M_{N}^{*}$ with increasing temperature 
resembles a phase transition at a high temperature and low density, when 
the system becomes a dilute gas of baryons in a sea of 
baryon-antibaryon pairs\cite{Furnstahl}.This behavior is consistent with 
the Walecka model\cite{Furnstahl} and resolves the
contradiction that appeared in the earlier calculations.
Below the critical temperature the effective mass grows 
with temperature. Above the critical temperature, $\sigma$
increases with temperature thus reducing 
the nucleon effective mass $M^{*}_{N}$. 

Finally,  we display the density dependence of the bag constant for different 
values of the temperature in Fig. 4.  
The bag constant as shown in Fig.4 (a), grows with temperature for temperatures 
less than 200 MeV (except at densities smaller than 0.1 fm$^{-3}$ where B starts 
to decrease for temperatures greater than 150 MeV). 
However, the situation  is completely reversed after the phase transition 
takes place. 
This  is displayed  in Fig.4 (b), where the bag constant displays a dramatic decrease 
with temperature for all densities at temperatures greater than 200 MeV. 
This indicates the onset of quark deconfinement above the critical temperature: 
at very high temperature and/or density the hadrons will dissolve into 
a quark-gluon plasma through what is  believed to be a first order phase transition. 
Because the QMC model, despite its limitations and shortcomings,  
uses the quark degrees of freedom explicitly, 
it has made it possible to observe the quark deconfiment phase transition 
which would not have been possible if only the nucleonic degrees of freedom are used. 
It remains to be seen if these calculations  can also be carried out  with 
more sophisticated nucleonic models and ultimately corroborated by QCD calculations.

%%%%%%%%%%%%%%%%
\acknowledgments
This work is supported by a grant from the Deutsche Forschungsgemeinscaft.
The authors dedicate this article to the memory of Prof. J. M. Eisenberg 
who was involved with early discussions about this work and whose sudden 
death has put an end to a promising collaboration.

%%%%%%%%%%%%%%%%%%%%%%%%

%%%%%%%%%%%%%%%%%%%%%%%%%%%%%%%%%%%%%%%%%%%%%%%%%%%%%%%%%%%%%%%%%%%%
%%\clearpage
% figures follow here
%

\clearpage

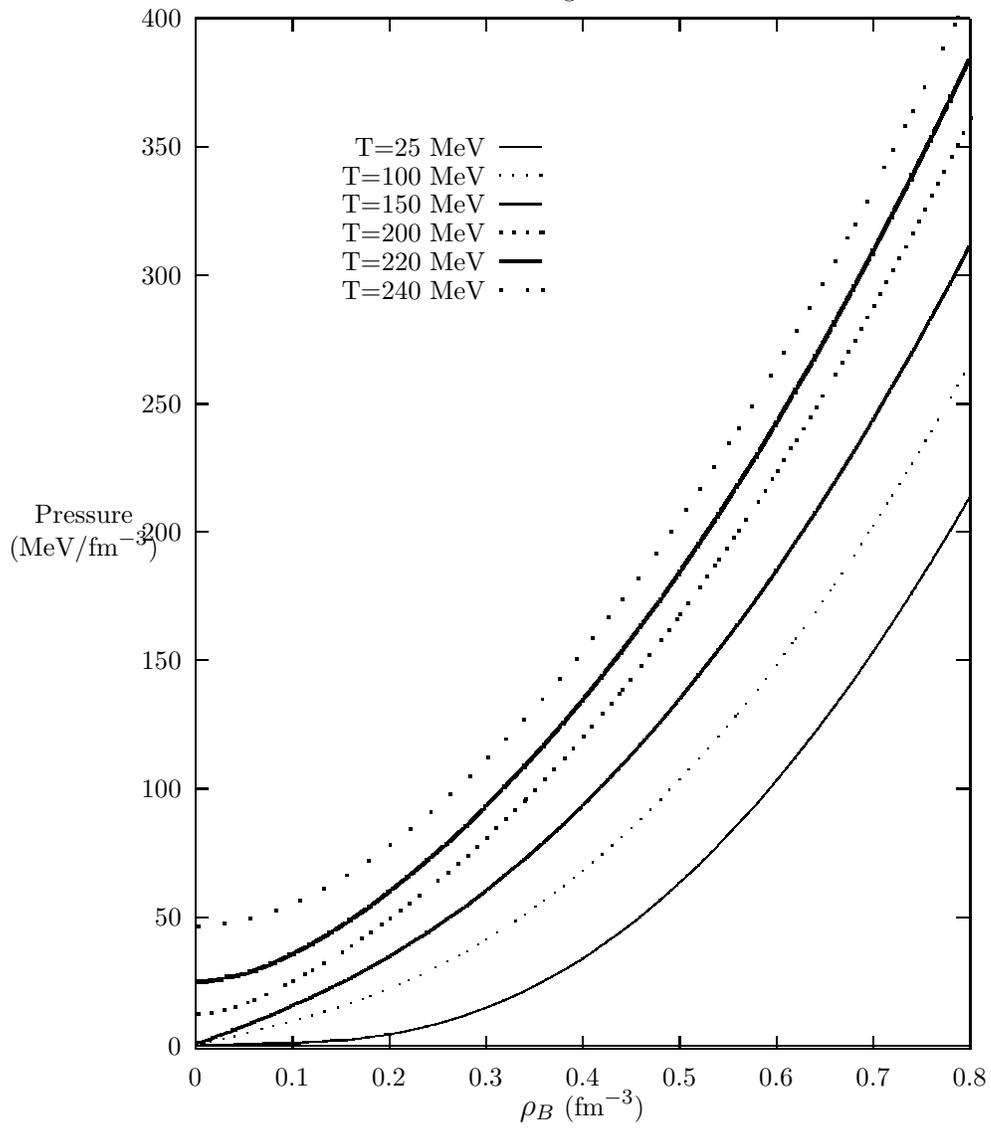
\begin{figure}
\caption{The pressure for nuclear matter as a function of 
the baryon density $\rho_{B}$ for various values of temperature.}
\label{f1}
\end{figure}

\begin{figure}
\caption{The mean scalar field $\sigma$ as a function 
of the baryon density $\rho_{B}$, 
(a) for different temperatures up to T=200 MeV,
(b) for different temperatures above T=200 MeV.}
\label{f2}
\end{figure}

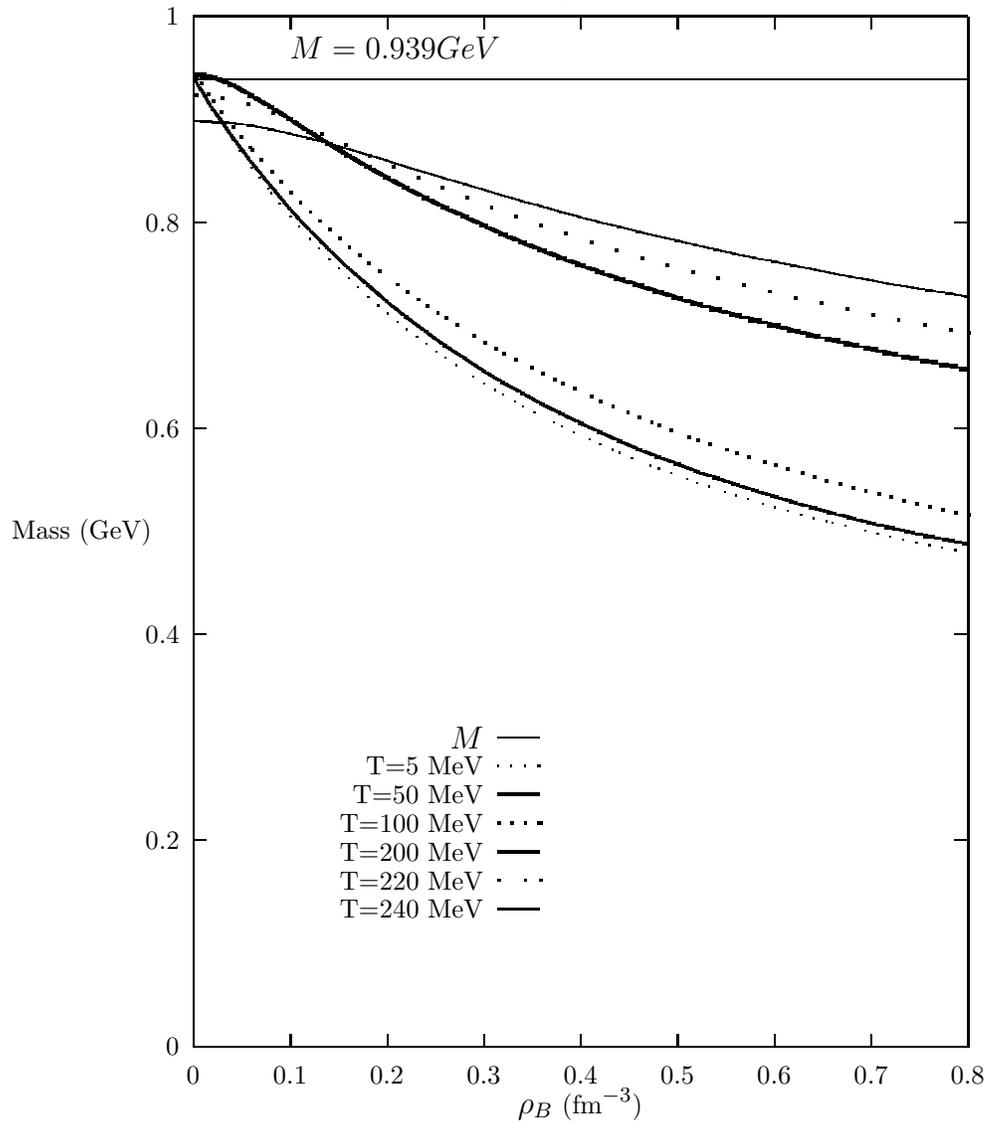
\begin{figure}
\caption{The effective nucleon mass $M_{N}^{*}$ 
for nuclear matter as a function of the baryon density 
$\rho_{B}$ for different values of temperature.}
\label{f3}
\end{figure}

\begin{figure}
\caption{The bag constant versus the baryon density $\rho_{B}$, 
(a) for different temperatures up to T=200 MeV, 
(b) for different temperatures above T=200 MeV.}
\label{f4}
\end{figure}

%
%

%
%\end{document}
%

\clearpage
\begin{figure}[htbp]
\begin{center}
\input{plotp1.tex}
\end{center}
\end{figure}
\clearpage  
\begin{figure}[htbp]
\begin{center}
\input{plotp2a.tex}
\end{center}
\end{figure}  
\clearpage
\begin{figure}[htbp]
\begin{center}
\input{plotp2b.tex}
\end{center}
\end{figure}
\clearpage
\begin{figure}[htbp]
\begin{center}
\input{plotp3.tex}
\end{center}
\end{figure}
\clearpage
\begin{figure}[htbp]
\begin{center}
\input{plotp4a.tex}
\end{center}
\end{figure}
\clearpage
\begin{figure}[htbp]
\begin{center}
\input{plotp4b.tex}
\end{center}
\end{figure}
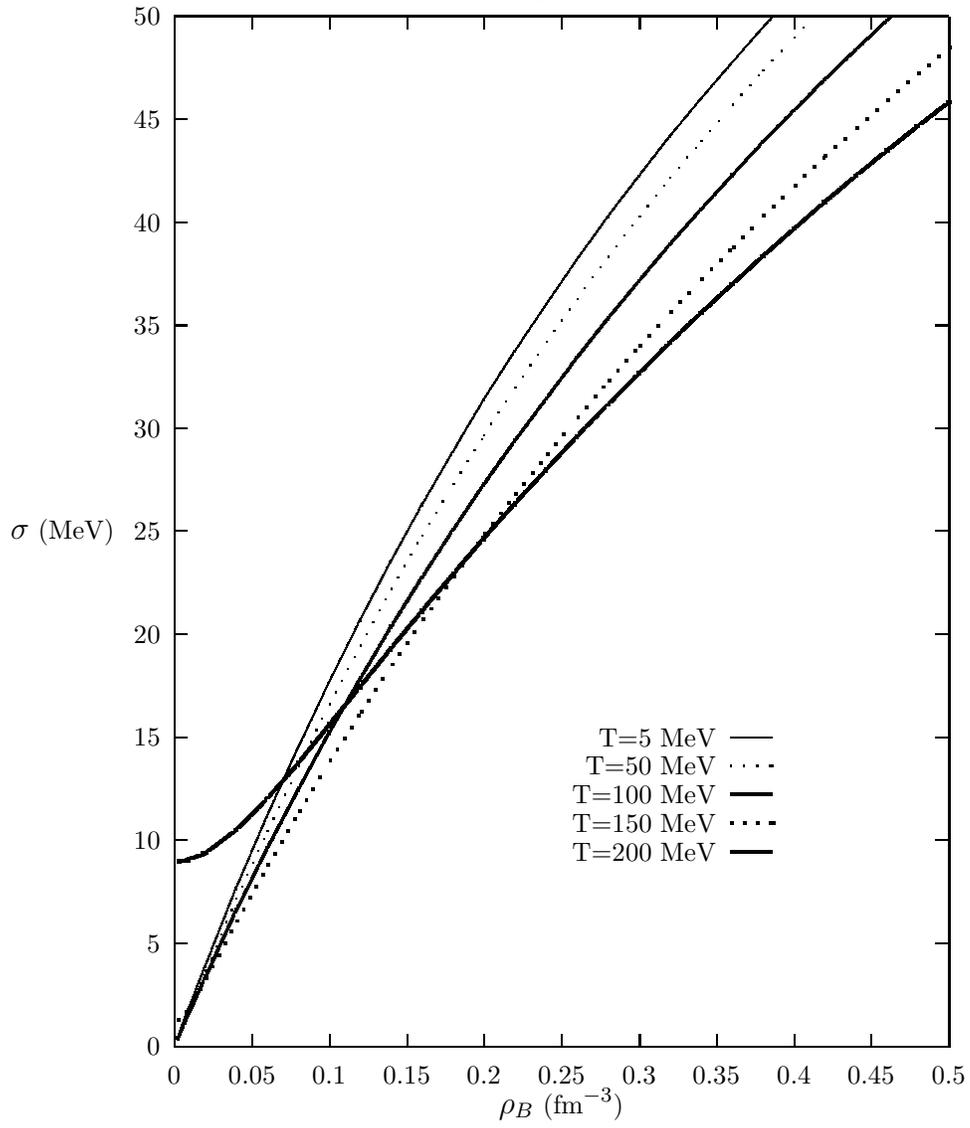
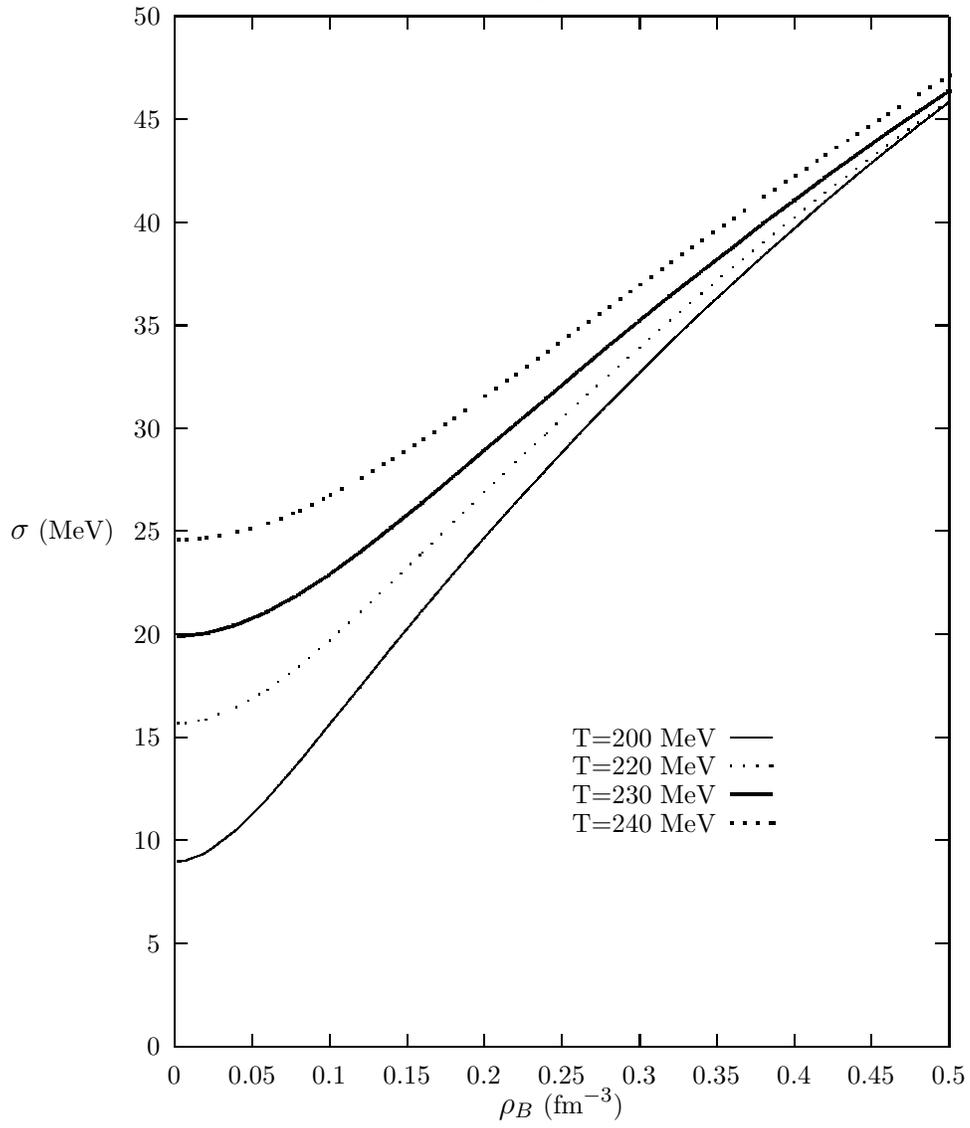
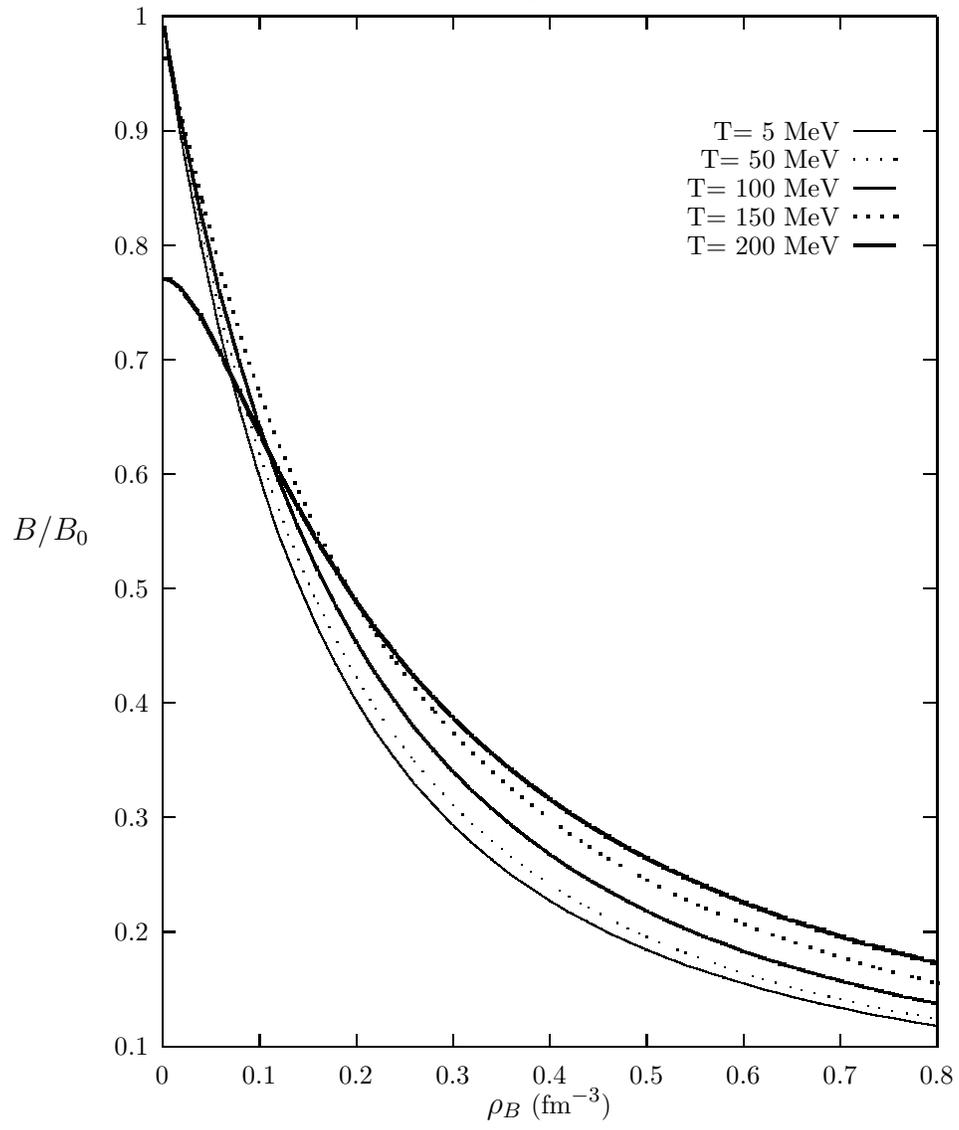
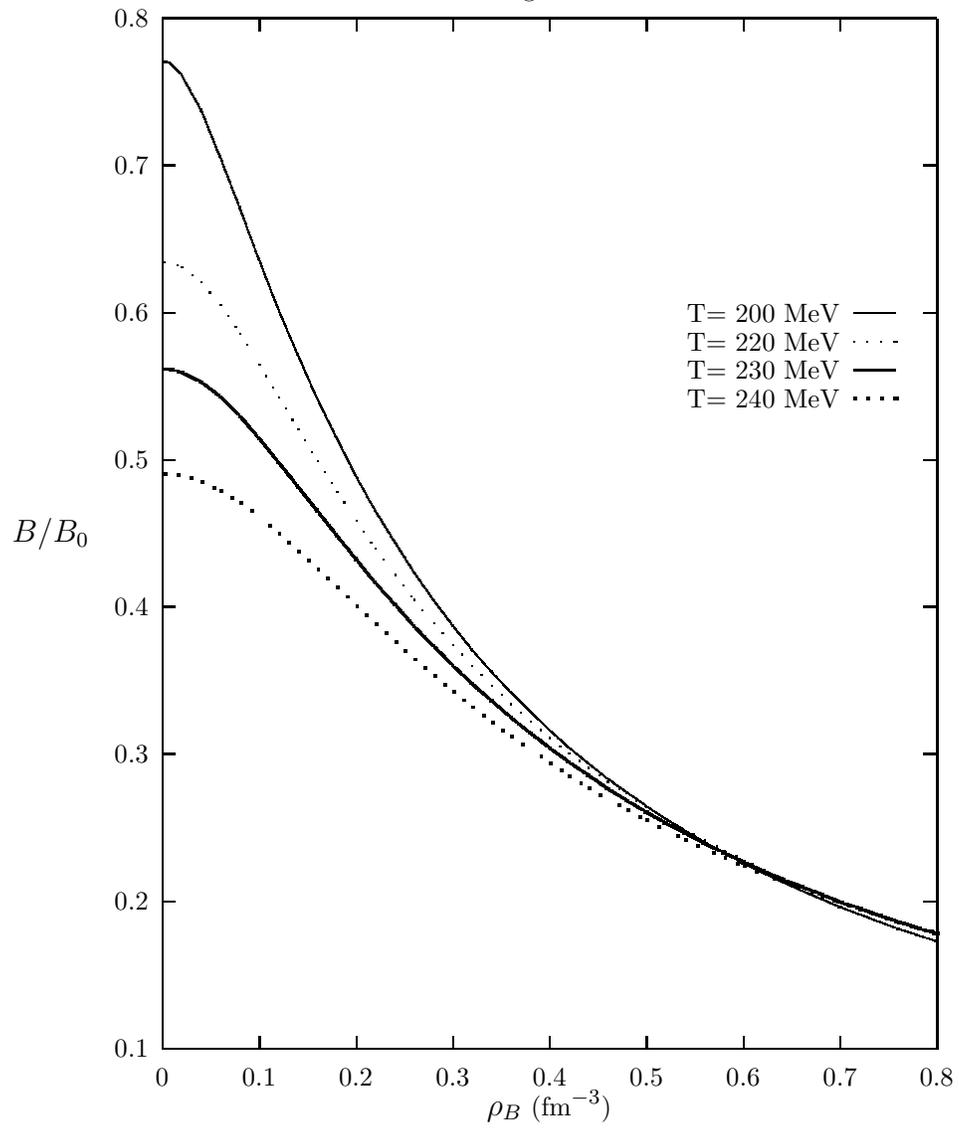
\end{document}